\documentclass[pre,twocolumn,fullpage,floatfix]{revtex4-1} 
\usepackage{graphicx,subfigure,color,amsmath,amssymb}

\begin{document}

\title{Extreme Thouless effect in a minimal model of dynamic social networks}
\author{K. E. Bassler$^{1,2,3}$, Wenjia Liu$^{4,5}$, B. Schmittmann$^{4}$, and
R. K. P. Zia$^{3,4,6}$}
\affiliation{$^1$ Department of Physics, University of Houston, Houston, TX 77204, USA \\
$^2$ Texas Center for Superconductivity, University of Houston, Houston, TX
77204, USA \\
$^3$ Max-Planck-Institut f\"{u}r Physik komplexer Systeme, N\"{o}thnitzer
Str. 38, Dresden D-01187, Germany \\
$^4$ Department of Physics and Astronomy, Iowa State University, Ames, IA
50011, USA \\
$^5$ Amazon-Blackfoot 1918 8th Ave, Seattle, WA 98101, USA \\
$^6$ Department of Physics, Virginia Polytechnic Institute and State
University, Blacksburg, VA 24061, USA}

\begin{abstract}
In common descriptions of phase transitions, first order transitions are
characterized by discontinuous jumps in the order parameter and normal
fluctuations, while second order transitions are associated with no jumps
and anomalous fluctuations. Outside this paradigm are systems exhibiting
`mixed order transitions' displaying a mixture of these characteristics. 
When the jump is maximal and the fluctuations range over the entire
range of allowed values, the behavior has been coined an `extreme Thouless
effect'. Here, we report findings of
such a phenomenon, in the context of dynamic, social networks. Defined by
minimal rules of evolution, it describes a population of extreme 
introverts and extroverts, who prefer to have
contacts with, respectively, no one or everyone. From the dynamics, we
derive an exact distribution of microstates in the stationary state. With
only two control parameters, $N_{I,E}$ (the number of each subgroup), we
study collective variables of interest, e.g., $X$, the total number of 
$I$-$E $ links and the degree distributions. Using simulations and mean-field
theory, we provide evidence that this system displays an extreme Thouless
effect. Specifically, the fraction $X/\left( N_{I}N_{E}\right) $ jumps from 
$0$ to $1$ (in the thermodynamic limit) when $N_{I}$ crosses $N_{E}$, while
all values appear with equal probability at $N_{I}=N_{E}$.
\end{abstract}


\maketitle


\section{Introduction}

In systems with many interacting degrees of freedom, interesting collective
phenomena are associated with phase transitions, e.g., in ferromagnetism.
Here, a suitable macroscopic variable characterizing the state of the system --
the order parameter -- typically changes its behavior in some rather
dramatic fashion. In standard
textbooks, phase transitions are classified by the Ehrenfest scheme: first
order, second order, etc. We also learn to expect certain characteristics
associated with each order. Thus, across the first order transition, the
order parameter jumps, while its fluctuations are `normal' (on either side).
Metastability, hysteresis, and co-existence are other common features
associated with this kind of transition. By contrast, opposite
characteristics, e.g., no discontinuity and anomalously large fluctuations,
are associated with second order transitions.

Though such properties are observed in most physical systems, there are
exceptions. In the context of one-dimensional Ising models with long range
interactions, Thouless \cite{Thouless,Aize88,Luij01} found `mixed order' transitions, at
which the order parameter jumps discontinuously and exhibits large
fluctuations. Since then, several systems with such properties have been
discovered \cite{Blossey95,Poland66,Fisher66,Kafri00,Gross85,Schwarz06,Toni06}. 
In particular, the term `extreme Thouless effect'
was coined recently \cite{BarMukamel14,BM14a} to describe a case where, at the
transition, both the jump and the fluctuations are maximal. In this paper,
we report another system displaying such an effect, in the context of a
minimal model of social interactions, involving dynamic networks with
preferred degrees.

In our previous studies of such networks \cite{LiuJoladSchZia13,LiuSchZia14},
we introduced the notion that an individual (i.e., node) adds/cuts links
to others according to its `preferred degree': $\kappa $. The evolution of
the simplest version of such networks is: In each time step, a random node
is chosen and its degree, $k$, is noted. If $k>\kappa $, the node cuts one
of its existing links at random. Otherwise, it adds a link to a randomly
chosen node not connected to it. In the steady state of a homogeneous system
(all nodes assigned the same $\kappa $), this ensemble of apparently random
graphs displays quite different properties \cite{CSP24} from the standard 
Erd\H{o}s-R\'{e}nyi case \cite{E-R}. Taking a small step towards describing 
an inhomogeneous society, we consider a heterogeneous system of two subgroups,
with $N_{1,2}$ nodes assigned different $\kappa $'s. Letting 
$\kappa_{1}<\kappa _{2}$, we naturally refer to the first group as `introverts' 
($I$) and the latter one as `extroverts' ($E$). Despite the simplicity of its
rules, such a system exhibits a rich variety of properties, discovered
mainly through simulations \cite{LiuSchZia14}. On the analytic front,
progress has been modest, since the underlying dynamics violates detailed
balance and the stationary state will have non-trivial persistent
probability currents \cite{ZS2007}. Under these circumstances, to gain some
insight, we turn to limiting cases which embody the main features of the
full system. In this spirit, we consider the ultimate limit: 
$\kappa_{1}=0,\kappa _{2}=\infty $. In other words, these are 
e\underline{x}treme \underline{i}ntroverts and \underline{e}xtroverts 
(or $XIE$, for short).
Specifically, we are able to find an exact expression for the stationary
distribution, obtain analytic predictions for various quantities which are
confirmed by simulations, and provide good evidence that the transition
across $N_{1}=N_{2}$ displays an extreme Thouless effect. While preliminary
results have been reported earlier \cite{CSP25,LiuSchmittmannZia12}, we will
present a more detailed study of this model here.

First, for the readers' convenience, we provide a summary of the preliminary 
findings, by including a complete description of the model, the master 
equation associated with the stochastic process, the exact microscopic 
distribution in the steady state, and a mapping to an two-dimensional Ising 
model with peculiar interactions. In Section III, we study the statistical 
properties of $X$ (the total number of I-E links), an `order parameter' 
which corresponds to the magnetisation of the Ising model. Using Monte Carlo 
techniques, we report findings much beyond those in ref. 
\cite{CSP25,LiuSchmittmannZia12}: including a power spectrum study of the 
time traces of X, as well as some first steps towards a finite size 
analysis for X, as a function of $N_{I,E}$ (the number of I,E's). These provide
further evidence for the principal characteristics associated with an extreme 
Thouless effect. The following section is devoted to investigations of a 
standard characterization of networks: degree distributions. In the Ising 
language, these correspond to novel measures of the system, offering a 
more detailed picture than the magnetisation. For systems with 
$N_{I}\neq N_{E}$, \textit{predictions} (i.e., no fitting parameters) 
of a self-consistent mean-field theory are largely confirmed by simulations. 
We end with a summary and outlook, while the Appendices contain many of 
the technical details.

\section{The $XIE$ model and the stationary distribution}

With the motivations for this model presented in both the introduction and
previous studies \cite{CSP25,LiuSchmittmannZia12}, we provide the
specifications of our model, using the language of a social network. Our
population consists of $N$ individuals, divided into two groups: $N_{I}$
introverts and $N_{E}$ extroverts. Their behavior is `extreme,' in the sense
that the former/latter prefers contacts to none/everyone. The rules of
evolution cannot be simpler:

\begin{itemize}
\item In each time step, a random individual is chosen.

\item If an introvert is chosen, it cuts a random existing link.

\item If an extrovert is picked, it adds a link to a random individual not
already connected to it.
\end{itemize}

\begin{flushleft}
Note that one link changes at every step, except when the chosen individual
is `content,' i.e., a totally isolated $I$ or a fully connected $E$.
Obviously, such a contented pair cannot be present simultaneously in our
model. In this sense, our system may be termed `maximally frustrated,' a
measure \cite{LiuSchZia14} we will not pursue here.
\end{flushleft}

In simulation studies, it is customary to define one Monte Carlo Step (MCS)
as $N$ such attempts, so that each node has an even chance of being chosen
after $N$ attempts. Our main interest will be the statistical properties of
this system at long times, once it settles into the stationary state. We
emphasize again that, in this minimal model, there are just two control
parameters: $N_{I,E}$. For large $N$, we may consider different
`thermodynamic' limits, e.g., fixed difference $N_{E}-N_{I}$ or ratio $%
N_{E}/N_{I}$. Note also that, though the total number of links can reach $%
N\left( N-1\right) /2$, the maximum number of \textit{cross-links} between
the two groups is%
\begin{equation}
\mathcal{N}\equiv N_{I}N_{E}
\end{equation}


\begin{figure}[tbp]
\centering
\includegraphics[width=2.8in]{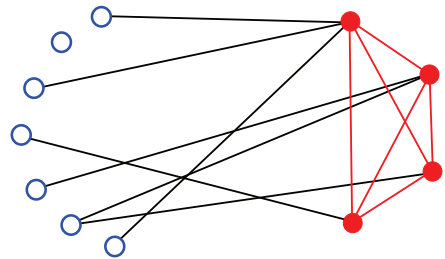}
\caption{(Color online) The nodes of the two groups are denoted by circles: 
blue open ($I$) and red closed ($E$). The black lines represent the active 
cross-links and the red dashed lines, the frozen $E$-$E$ links. For this 
network, the sets of $k$'s are: $k_{I}=\left\{ 1,0,1,1,1,2,1\right\} $, 
and $k_{E}=\left\{6,5,4,4\right\} $. Thus, this configuration contributes 
$1,5,1$ to $\protect\rho _{I}$ ($k=0,1,2$) and $2,1,1$ to 
$\protect\rho _{E}$ ($k=4,5,6$), respectively.}
\label{nodes}
\end{figure}


Clearly, regardless of how the system is initialized, all the
intra-communities links ($I$-$I$ or $E$-$E$) will quickly become static (all
absent or present). Only the $I$-$E$ cross-links are dynamic, depending on
which node happens to be chosen (illustrated in Fig. \ref{nodes}). In other
words, we may limit our attention to the set of bipartite graphs. Of course,
the total number of cross-links, $X$ ($\in \left[ 0,\mathcal{N}\right] $),
fluctuates and should display interesting statistical properties. The first
impression of such a minimal model is that it must be trivial. In
particular, it may be argued that, since the probability for a link to be
cut or added is proportional to $N_{I}/N$ or $N_{E}/N$, the fraction 
\begin{equation}
x\equiv X/\mathcal{N}
\end{equation}%
should be simply $N_{E}/N$. In the rest of this article, we will show the
behavior of $x$ to be dramatically different.

The configuration space here consists of the $N_{I}\times N_{E}$ incidence
matrices, $\mathbb{N}$. We denote its elements by $n_{ij}$ ($i\in \left[
1,N_{I}\right] ,j\in \left[ 1,N_{E}\right] $) which is $1\,$or $0$ when the
link between an introvert node $i$ and an extrovert $j$ is present or
absent, respectively. Clearly, this configuration space is identical to that
of a $N_{I}\times N_{E}$ Ising model on a square lattice. Following the
lattice gas language \cite{YangLee52}, we will refer to $n=1,0$ as a
`particle' or a `hole.' Meanwhile, the dynamics of cutting/adding
corresponds to a kinetic Ising model with spin flip dynamics \cite{Glauber63}
(i.e., without particle conservation). Since $X=\Sigma _{ij}n_{ij}$ and $%
\mathcal{N}$ is the total number of sites in the Ising system, we have the
mapping 
\begin{equation}
x=\left( 1+m\right) /2
\end{equation}%
where $m\in \left[ -1,1\right] $ is the magnetisation. To further this
correspondence, let us define%
\begin{equation}
h\equiv \Delta /N;~~\Delta \equiv N_{E}-N_{I}  \label{h+Delta}
\end{equation}%
with $h$ playing the role of the magnetic field. Thus, our interest in how $%
x $ responds to $N_{I,E}$ translates into finding a `equation of state' $%
m\left( h\right) $. In this language, the naive expectation is trivial (and
false): $m\left( h\right) =h$.

Of course, unlike in the Ising case, the statistical mechanics of the $XIE$
model is defined by dynamical rules rather than a Hamiltonian. Therefore, to
study quantities of interest, we must first find the distribution of the
stationary state, $\mathcal{P}^{ss}\left( \mathbb{N}\right) $, as opposed to
just writing a Boltzmann factor.

With the dynamics specified, we can write the master equation for $\mathcal{P%
}(\mathbb{N},t)$, the probability of finding the system in configuration $%
\mathbb{N}$ at time $t$: 
\begin{eqnarray}
&&\mathcal{P}(\mathbb{N},t+1)-\mathcal{P}(\mathbb{N},t)=  \notag \\
&&\qquad \sum_{\{\mathbb{N}^{\prime }\}}[W(\mathbb{N},\mathbb{N}^{\prime })%
\mathcal{P}(\mathbb{N}^{\prime },t)-W(\mathbb{N}^{\prime },\mathbb{N})%
\mathcal{P}(\mathbb{N},t)]  \label{ME}
\end{eqnarray}%
where $W(\mathbb{N}^{\prime },\mathbb{N})$ is the probability for
configuration $\mathbb{N}$ to become $\mathbb{N}^{\prime }$ in a step (an
attempt). For $W$, we note that the dynamics for an $I$ node $i$ involves
cutting a random \textit{existing}\textbf{\ }link, so that $k_{i}$, the
number of links it has (i.e., particles in row $i$ in $\mathbb{N}$) will be
needed. Similarly, for a $E$ node $j$, we will need $p_{j}$, the number of
holes in column $j$ in $\mathbb{N}$. Letting $\bar{n}_{ij}\equiv 1-n_{ij}$,
we define%
\begin{equation}
k_{i}\equiv \Sigma _{j}n_{ij};~~p_{j}\equiv \Sigma _{i}\bar{n}_{ij}
\label{kpbar}
\end{equation}%
Clearly, these variables reveal a not-so-explicit symmetry in the dynamics
of $XIE$. Similar to the Ising spin flip symmetry ($n\Leftrightarrow \bar{n}$%
), there is an additional, transpose operation:%
\begin{equation}
n_{ij}\Leftrightarrow \bar{n}_{ji}\oplus N_{I}\Leftrightarrow N_{E}
\label{ph-sym}
\end{equation}%
We will refer to this, \`{a} la Ising, as `particle-hole symmetry,' which
will play an important role in discussions below. A layman's way to phrase
this symmetry is: The presence of a link is as intolerable to an introvert
as the presence of a `hole' is to an extrovert.

With these preliminaries, the $W(\mathbb{N}^{\prime },\mathbb{N})$ in Eqn. 
(\ref{ME}) reads%
\begin{equation}
\sum\limits_{i,j}\left[ \frac{\Theta \left( k_{i}\right) }{k_{i}}\bar{n}%
_{ij}^{\prime }n_{ij}+\frac{\Theta \left( p_{j}\right) }{p_{j}}%
n_{ij}^{\prime }\bar{n}_{ij}\right] \frac{\Pi _{k\ell \neq ij}\delta \left(
n_{k\ell }^{\prime },n_{k\ell }\right) }{N}  \label{rates}
\end{equation}%
where $\Theta \left( x\right) $ is the Heavyside function (i.e., $1$ if $x>0$
and $0$ if $x\leq 0$) and the product of $\delta $'s ensures that only one 
$n_{ij}$ may change in a step. It is straightforward to verify that Eqn. 
(\ref{ME}) respects particle-hole symmetry.

The dynamics defined by Eqn. (\ref{ME}) is clearly ergodic. More remarkably,
unlike those in less extreme models of introverts and extroverts \cite%
{LiuJoladSchZia13,LiuSchZia14}, it obeys detailed balance (shown in Appendix
A). Consequently, in the $t\rightarrow \infty $ limit, $\mathcal{P}$
approaches a unique stationary distribution, which can be found by applying $%
\mathcal{P}^{ss}\left( \mathbb{N}\right) =\mathcal{P}^{ss}\left( \mathbb{N}%
^{\prime }\right) W\left( \mathbb{N},\mathbb{N}^{\prime }\right) /W\left( 
\mathbb{N}^{\prime },\mathbb{N}\right) $ repeatedly. Imposing normalization,
we arrive at an explicit, closed form: 
\begin{equation}
\mathcal{P}^{ss}\left( \mathbb{N}\right) =\frac{1}{\Omega }%
\prod\limits_{i=1}^{N_{I}}\left( k_{i}!\right)
\prod\limits_{j=1}^{N_{E}}\left( p_{j}!\right)  \label{P*}
\end{equation}%
where $\Omega =\Sigma _{\left\{ \mathbb{N}\right\} }\Pi \left( k_{i}!\right)
\Pi \left( p_{j}!\right) $ is a `partition function.' Note that the
particle-hole symmetry (Eq.~\ref{ph-sym}) is manifest here.

Interpreting $\mathcal{P}^{ss}$ as a Boltzmann factor and trivially assuming 
$\beta=1$, we can write a `Hamiltonian' \footnote{%
Of course, $\ln \left( \Sigma _{j}n_{ij}\right) !$ can be cast as $\Sigma
_{\ell }\ln \left( \Sigma _{j}n_{ij}-\ell \right) $ but this form is hardly
a simplification.} 
\begin{equation}
\mathcal{H}\left( \mathbb{N}\right) =-\left\{ \sum\limits_{i=1}^{N_{I}}\ln
\left( \sum\limits_{j=1}^{N_{E}}n_{ij}\right) !+\sum\limits_{j=1}^{N_{E}}\ln
\left( \sum\limits_{i=1}^{N_{I}}\bar{n}_{ij}\right) !\right\}  \label{H}
\end{equation}%
Now, this expression immediately alerts us to the level of complexity of
this system of `Ising spins,' as $\mathcal{H}$ contains a peculiar form of
long range interactions. Each `spin' is coupled to all other `spins' \textit{%
in its row and column}, via all possible types of `multi-spin' interactions!
We are not aware of any system in solid state physics with this kind of
interactions. It is remarkable that such a complex $\mathcal{H}$ emerges
from an extremely simple model of social interactions. Meanwhile, it is
understandable that computing $\Omega $, let alone statistical properties of
macroscopic quantities, will be quite challenging (Appendix B).
Nevertheless, as the next two sections show, we are able to exploit
mean-field approaches to gain some insight, and to \textit{predict }some
macroscopic observables. For generic $\left( N_{I},N_{E}\right) $, we find
excellent agreement with simulation data. As in standard equilibrium
statistical systems, these theories fail in the neighborhood of critical
points, which turn out to be the $N_{I}=N_{E}$ line here.

\section{Statistics of $X$, the total number of cross-links}

In the $XIE$ model, the most natural macroscopic quantity to study is $X$,
the total number of $I$-$E$ links. In addition, its average $\left\langle
X\right\rangle $ can serve as an `order parameter,' since it plays the role
of the total magnetisation in the Ising model. Though there is no natural
temperature-like variable in our model of social networks, there is a
natural external-field-like control parameter: $h$ (or $\Delta $). Our main
interest in this section is how $\left\langle X\right\rangle $ varies with 
$N_{I,E}$. In other words, what is the `equation of state' 
$m\left( h;\mathcal{N}\right) $ for the $XIE$ model? If it were like the Ising model below criticality, 
$m\left( h=0_{\pm };\mathcal{N}\rightarrow \infty \right)=\pm m_{s}$, where 
$0<m_{s}<1$ is the spontaneous magnetisation, accompanied
by ordinary $O\left( 1/\sqrt{\mathcal{N}}\right) $ fluctuations. In the
Thouless effect, a discontinuity would be accompanied by anomalously large
fluctuations. Here, we will find that our $m$ displays an \textit{extreme}
Thouless effect in the thermodynamic limit, i.e., 
$m\left( h=0_{\pm };\mathcal{N}\rightarrow \infty \right) =\pm 1$, 
together with extraordinary, $O\left( 1\right) $, fluctuations around 
$m\left( h=0\right) =0$.

\begin{figure}[tbp]
\centering
\includegraphics[width=3.0in]{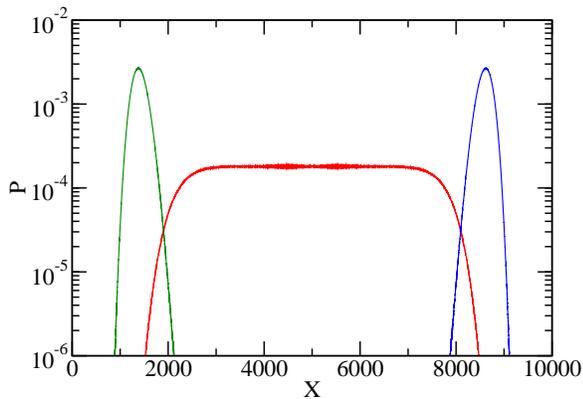}
\caption{(Color online) The distributions, $P\left( X\right)$, 
compiled with the time traces of $X$ for three cases with 
$\left( N_{I},N_{E}\right) $ near and at criticality: 
$\left( 101,99\right)$ (green dashed line), 
$\left( 99,101\right)$ (blue dot-dashed line), 
and $\left(100,100\right)$ (red solid line).}
\label{histogram}
\end{figure}

The preliminary results for $N=200$, displayed in Figs. 1-3 in \cite%
{LiuSchmittmannZia12} gave a hint of these remarkable properties. To confirm
and to improve on those results, we carry out much longer runs: discarding
the first $5\times 10^{7}$ MCS, taking measurements every $50$ MCS, for up
to $10^{11}$ MCS (for the $N_{I,E}=100$ case). Using these long traces, 
$X\left( t\right) $, we find much more information than just the average 
$\left\langle X\right\rangle $; we obtain a much more accurate picture for
the whole steady state distribution $P\left( X\right) $: Fig.~\ref{histogram}%
. Not surprisingly, they are sharply peaked and Gaussian-like for the
off-critical cases (green and blue on line), while the distribution in the 
$N_{I}=N_{E}$ case (red on line) is essentially flat over most of the full
range, $\left[ 0,\mathcal{N}\right] $. The flat plateau in $P\left( X\right) 
$ gives the impression of an \textit{unbiased} random walk (bounded by `soft'
walls near the extremes of the allowed region). In stark contrast,
characteristic of co-existence in an ordinary first order transition, 
$M\left( t\right) $ for an Ising system below criticality spends much of its
time hovering around the spontaneous magnetizations, $\pm M_{sp}$, and makes
rare and short excursions from one to the other.


\begin{figure}[tbp]
\centering
\includegraphics[width=3.5in]{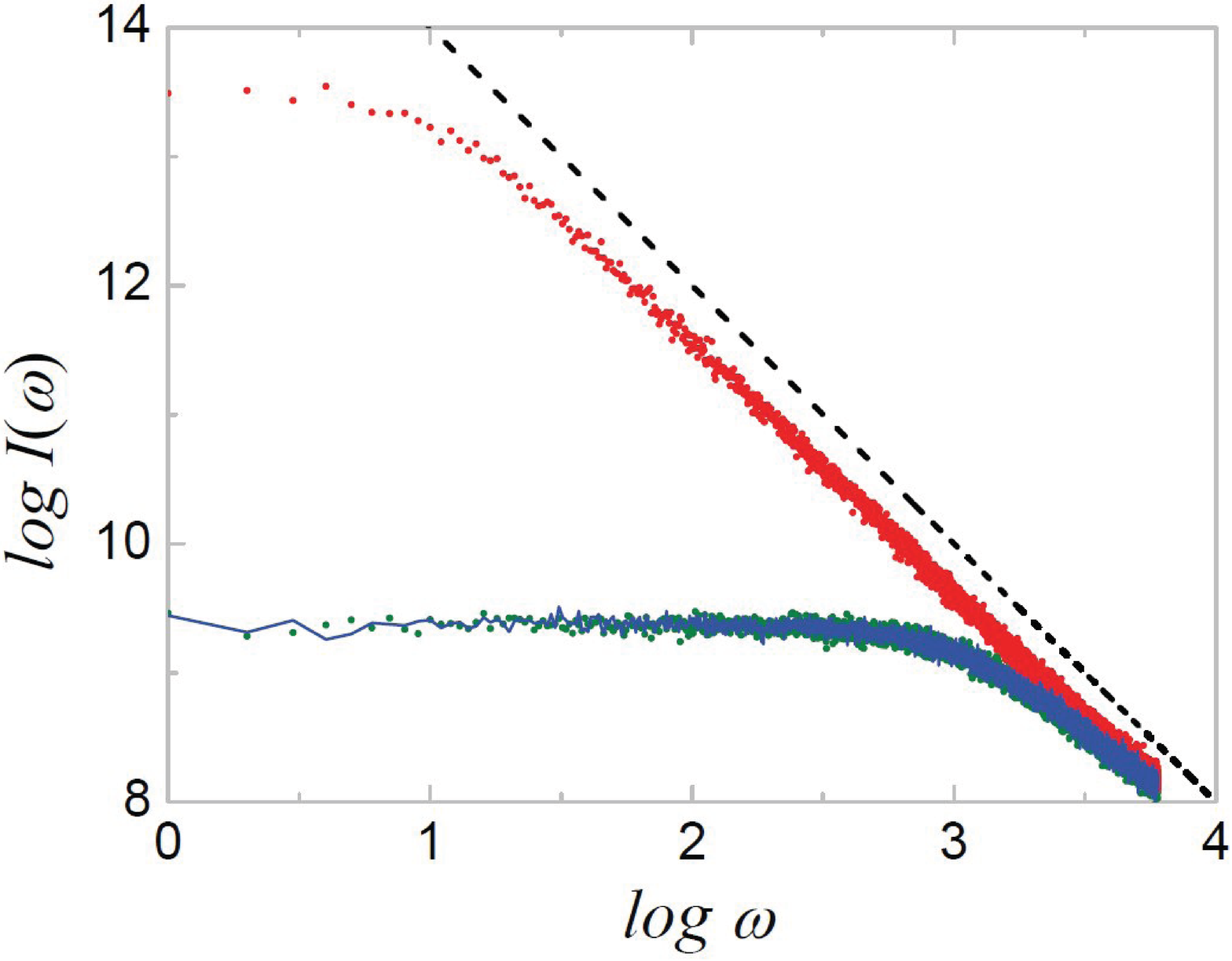}
\caption{(Color online) Power spectra, $I\left( \protect\omega \right) $, 
associated with the time traces of $X(t)$. The dashed black line is
proportional to $1/\protect\omega ^{2}$. Note that the spectra associated
with the two off-critical cases (farthest from the dashed line, shown 
as green dots and a blue line) are statistically identical, as expected 
from particle-hole symmetry. Both scales are $log_{10}$.}
\label{PS}
\end{figure}

Another principal characteristic is metastability or hysteresis. Though not
shown explicitly here, we observe neither. When the two nodes `change
sides,' i.e., $\left( 101,99\right) \rightarrow \left( 99,101\right) $, $X/%
\mathcal{N}$ simply marches from $\thicksim 15\%$ to $\thicksim 85\%$ in $%
\thicksim 3500$ MCS. In other words, on the average, $X$ changes by about
two links per MCS. We also considered having $\delta =4$ or $6$ `defectors'
instead of just two, in systems with $N=400$ and $800$. In all cases, the
average `velocity' is approximately $\delta $ per MCS. Intuitively, we may
attribute this to the action of the $\delta $ extra $E$ nodes, but it
remains to be shown analytically. In all respects, there is absolutely 
\textit{no barrier} between the two extremes of $X$~! As for criticality
itself $\left( 100,100\right) $, the notion of $X$ executing a pure random
walk (RW) can be further confirmed by studying its power spectrum. With $%
\mathcal{T}=2\times 10^{4}$ measurements (of runs of $2\times 10^{6}$ MCS),
we compute the Fourier transform $X\left( \omega \right) $ and then average
over 100 runs to obtain $I\left( \omega \right) \equiv \left\langle
\left\vert X\left( \omega \right) \right\vert ^{2}\right\rangle $. In Fig.~%
\ref{PS}, we show plots of $\log I$ \textit{vs.} $\log \omega $, as well a
straight line (black dashed) representing $\omega ^{-2}$. The upper set of
data points (red on line), associated with $N_{I}=N_{E}=100$, are
statistically consistent with the RW characteristic of $\omega ^{-2}$. The
cutoff at small $\omega $ can be estimated from the finite range available
to the RW ($\thicksim 7000$ here). Since $\Delta X=\pm 1$ in each attempt,
we can assume the traverse time to be about $7000^{2}\cong 5\times 10^{7}$
attempts, or $\thicksim 2.5\times 10^{5}$ MCS. Given that this value is
comparable to $1/10$ of our run time, it is reasonable to expect deviations
from the pure $\omega ^{-2}$ as we approach $\omega \thicksim 10$. By
contrast, the power spectra of the two off-critical cases (lower set of
data, green and blue on line) are controlled by some intrinsic time scale
associated with both the restoration to $\left\langle X\right\rangle $ and
the fluctuations thereabout. Indeed, this $I\left( \omega \right) $ is
entirely consistent with a Lorentzian, i.e., $\propto 1/\left( \omega
^{2}+\omega _{0}^{2}\right) $. Given our limited understanding of the
dynamics of this model, estimating $\omega _{0}$ is beyond the scope of this
work.


As shown in ref. \cite{LiuSchmittmannZia12}, some characteristics of an
extreme Thouless effect (presence of a jump, absence of metastability and
hysteresis, existence of a flat plateau in $P\left( X\right) $, etc.) can be
qualitatively understood in terms of a crude mean field approximation. That
approach starts with the exact%
\begin{equation}
P\left( X\right) \equiv \sum_{\left\{ \mathbb{N}\right\} }\delta \left(
X,\Sigma _{ij}n_{ij}\right) \mathcal{P}^{ss}\left( \mathbb{N}\right) .
\label{PX}
\end{equation}%
and replaces every $n_{ij}$ by its average $X/\mathcal{N}$ in the sum,
resulting in an approximate $\tilde{P}\left( X\right) $. Its maximum can be
used to predict $\left\langle X\right\rangle $ and so, an approximate
equation of state $\tilde{m}\left( h;\mathcal{N}\right) $. 
Remarkably, at the lowest order in $1/\mathcal{N}$, this approach leads 
to an \textit{extreme} Thouless effect \cite{BarMukamel14,BM14a}, i.e.,%
\begin{equation}
\tilde{m}=sign\left( h\right) ,
\end{equation}%
the absence of metastability and hysteresis when $h$ changes sign, as well
as a flat plateau in $\tilde{P}\left( X\right) $ for $h=0$! Keeping the next
order in $\tilde{F}$, we find%
\begin{equation}
\tilde{m}\left( h;N\right) =sign\left( h\right) -\frac{1}{hN}+...
\label{m(h)-mf}
\end{equation}%
which provides \textit{qualitatively} good agreement with most of the $N=200$
data set. Clearly, this mean-field approach captures some key features of
the $XIE$ model, even though it is not quantitatively reliable.

Before we turn to a much better mean field theory, let us present the
rudiments of a more complete portrait, towards a systematic scaling study.
Specifically, we study $m\left( h;N\right) $ in the transition region,
using simulations with $N$ up to $3200$. Of course, unlike the Ising model,
there is no natural variable in the social network corresponding to
temperature. Nevertheless, the results suggest the presence of anomalous
power laws and possible data collapse.

\begin{figure}[tbp]
\centering
\includegraphics[width=3.5in]{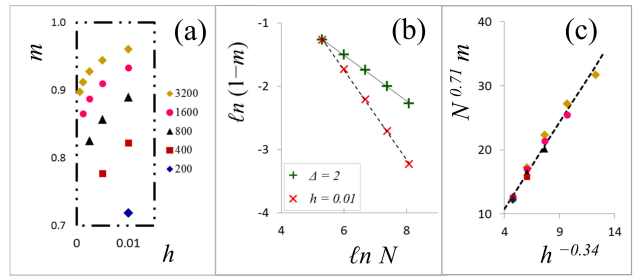}
\caption{(Color online) (a) $m(h)$ for various $N$'s and allowed 
$\Delta $'s with $h\leq 0.01$. (b) Log-log plots of indicated
subset of points in (a). Solid and dashed lines are linear fits with slope 
$-0.363$ and $-0.707$, respectively. Both correlation coefficients are
greater than $0.999$. (c) Scaled plots of the points in (a), showing
tolerable data collapse. Dashed line is linear, as guide to the eye.}
\label{ManyM}
\end{figure}

To study the behavior near criticality, we measure the average $\left\langle
X\right\rangle $ for all possible values of $N_{I,E}$ which correspond to $%
h\leq 0.01$, with $N=200,400,...,3200$. In other words, we use the
appropriate values of $\Delta =2,4,...32$. Starting with a half filled set
of $I$-$E$ links, each point was calculated with a run of 
up to $5\times 10^{9}$ MCS with measurements made about every $N/4$ MCS.
Verifying that our data is consistent with particle-hole symmetry, we present 
results only for $h>0$ and $m\cong 1$. Fig. \ref{ManyM}a shows results within 
the small regime $h\in \left[ 0,0.01\right] $ and $m\in \left[ 0.7,1.0\right] $. 
We see that, indeed, $m\left( h;\mathcal{N} \right) $ approaches $+1$ as 
$N\rightarrow \infty $. Note that, we can access smaller $h$ in 
systems with larger $N$, as its minimum is $2/N$
\footnote{With odd $N$, the smaller $h=1/N$ can be accessed. 
However, $N_{I}=N_{E}$ systems are not available.} .
With this set of data, we make two log-log plots (Fig. \ref{ManyM}b) 
showing that (i) for fixed $h=0.01$, $m\rightarrow 1$
with $N^{-0.71}$ and (ii) for fixed $\Delta =2$, $m\rightarrow 1$ with $%
N^{-0.36}$. The solid lines represent linear fits with correlation 
coefficient (R$^{2}$) greater than 0.999. 
Using this information, we plot $N^{0.71}m$ against $h^{-0.34}$ in
Fig. \ref{ManyM}c. Though somewhat rough, this plot does qualitatively
indicate data collapse and provides the first steps towards an in-depth
finite size scaling analysis. Such a study is beyond the scope of this paper
and will be reported elsewhere \cite{FGKB}. If the exponents found here are
confirmed, they signal a significant deviation from the mean-field values
(e.g., Eq.\ref{m(h)-mf}): $\left( 0.70,0.36\right) $ as opposed to $\left(
1,0\right) $. Of course, they also provide fertile ground for
renormalization group analysis, along the lines of ref. 
\cite{BarMukamel14, BM14a}.


\section{Degree distributions: simulation results and a dynamic mean-field
theory}

In this section, we turn to a `mesoscopic' quantity which offers a more
detailed perspective than the macroscopic $X$, as well as major contrasts
between Ising-like statistical systems and those associated with networks.
For the Ising model, a natural quantity to study is the total
magnetisation, which corresponds to $X$. But, it is not usual
to study the statistics of the magnetisation in a row or a column. Yet, for
the $XIE$ model, the corresponding quantity is the degree distribution, $%
\rho (k)$, which is one of the most common ways to characterise a network.
Thus, we devote the rest of this paper to these distributions, illustrating
with simulation data (for $N=200$), as well as offering a more effective
mean-field theory. Specifically, unlike the mean-field approach for $\tilde{P%
}\left( X\right) $, we will formulate an approximate \textit{dynamics for }$%
\rho $ and arrive at much better agreement with data, for \textit{all} $%
N_{I}\neq N_{E}$. In particular, $\tilde{m}\left( 0.01;200\right) $ above
differs from data by $\thicksim 15\%$, while the theory below produces a
value within 0.02\%.

Before presenting the data, let us set the stage for discussing \textit{two }%
degree distributions. In general, associated with a network with several
subgroups or communities, we can study many such distributions, to describe
the various intra- and inter-community links. For $XIE$, the intra-community
links are static and so, we need to study only two: $\rho _{I}(k_{I})$ and $%
\rho _{E}(k_{E})$, related to the degrees of the $I$'s and $E$'s,
respectively. To illustrate, a network with $\rho _{I}(1)=5$ and $\rho
_{E}(6)=1$ is shown in Fig.~\ref{nodes}. From the $\rho $'s, average degrees 
$\left\langle k_{I}\right\rangle $ and $\left\langle k_{E}\right\rangle $
can be found. Note that $\left\langle k_{I}\right\rangle \neq \left\langle
k_{E}\right\rangle $ typically, but they are related, in the steady state,
by the following. Since $N_{I}\left\langle k_{I}\right\rangle $ is just the
average number of cross-links, while an extrovert has $k_{E}-N_{E}+1$ links
to the $I$'s, we have%
\begin{equation}
N_{I}\left\langle k_{I}\right\rangle =\left\langle X\right\rangle
=N_{E}\left\langle k_{E}\right\rangle -N_{E}\left( N_{E}-1\right)
\label{constraint}
\end{equation}%
This complication can be bypassed, especially in view of the particle-hole
symmetry discussed above, by introducing a `hole' distribution for the $E$%
's: $\zeta _{E}\left( p_{E}\right) $. Clearly, $\zeta $ is intimately
related to $\rho $, namely, $\zeta \left( p\right) =\rho \left( N-1-p\right) 
$. Meanwhile, $N_{E}\left\langle p_{E}\right\rangle =\mathcal{N}%
-\left\langle X\right\rangle $, so that a manifestly symmetric constraint on 
$\rho ,\zeta $ is $N_{I}\left\langle k_{I}\right\rangle +N_{E}\left\langle
p_{E}\right\rangle =\mathcal{N}$.

\begin{figure*}[tbp]
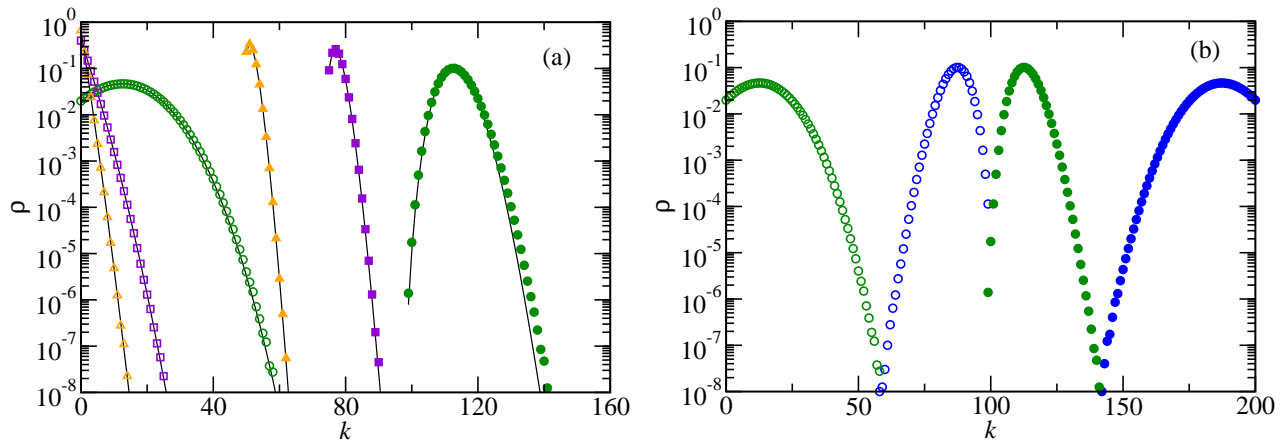

\centering
\mbox{
\subfigure{\includegraphics[width=3.25in]{DDa.eps}}\quad
    \subfigure{\includegraphics[width=3.25in]{DDb.eps}}
     }
\caption{(Color online) Degree distributions, $\protect\rho $, for several cases 
with $N_{I}+N_{E}=200$. Simulation results for the low/high $k$ components,
associated with introverts/extroverts, are denoted by open/solid symbols.
(a) The symbols for $(N_{I},N_{E})$ are orange triangles $(150,50)$, purple
squares $(125,75)$, and green circles $(101,99)$. The solid black lines are
predictions from a self-consistent mean-field theory. (b) When two
introverts `change sides,' a dramatic jump in $\protect\rho \left( k\right) $
results, with the case of $(99,101)$ shown as blue diamonds.}
\label{DD}
\end{figure*}

Next, let us present the simulation results. Starting with a network with
various initial conditions (null graph, complete graph, random half-filled),
we evolve the system according to the simple rules given above. Not
surprisingly, after $O\left( N\right) $ MCS, all the $I$-$I$ links are
absent while all the $E$-$E$ links are present. To be quite certain that the
system has equilibrated, we discard the first $5\times 10^{7}$ MCS.
Thereafter, we measure the degrees of each node every $50$ MCS. The
distributions are then obtained as the average over $10^{8}$ measurements.
Shown in Fig.~\ref{DD}(a) are $\rho \left( k\right) $ for three cases: $%
\left( N_{I},N_{E}\right) =\left( 150,50\right) $, $\left( 125,75\right) $, $%
\left( 101,99\right) $. Evidently, each $\rho $ consists of two components,
associated with $k\leq N_{E}$ and $k\geq N_{E}-1$. Generally, these
component are disjoint and so, they can be identified unambiguously with
respectively, $\rho _{I}$ and $\rho _{E}$ (shown with open and solid
symbols). Note that, in the $\left( 101,99\right) $ case, despite having
just two members less, the $\thicksim 100$ extroverts are unable to create
enough cross-links, so that there are essentially no $I$'s with say, $50$
links! Of course, this result is entirely consistent with our observations
above, showing that a typical $I$ has only about $15$ links. Apart from
having these two components, the most prominent feature is that neither
component resembles $3^{-\left\vert k-\kappa \right\vert }$, the degree
distribution of a homogeneous population with preferred degree $\kappa $ 
\cite{CSP24,LiuJoladSchZia13}. As will be shown, they are well approximated
by Poisson distributions, an analytic result of our mean-field theory.


What happens when the introverts `defect'? Though the changes appear
dramatic, they should not be a surprise, given the underlying particle-hole
symmetry in $XIE$. Thus, we illustrate in Fig.~\ref{DD}(b) the degree
distribution for $(N_{I},N_{E})=(99,101)$ (blue circles), as well as the
previous case of $(101,99)$ (green circles). By exchanging $%
N_{I}\Leftrightarrow N_{E}$ and plotting the degree distribution \textit{vs}%
. $p\equiv N-1-k$, we find perfect (within statistical errors) overlap
between the blue and green data points.

\subsection{Self-consistent mean-field approximation (SCMF)}

Given the exact steady state distribution (Eq.~\ref{P*}), the $\rho $'s can
be computed, in principle, from e.g., $\rho _{I}\left( k_{I}\right)
=\sum_{\left\{ \mathbb{N}\right\} }\delta \left( k_{I}-\Sigma
_{j}n_{ij}\right) \mathcal{P}^{ss}\left( \mathbb{N}\right) $ (for any $i$).
In practice, this task is as difficult as computing $P\left( X\right) $, so
that we again resort to a mean-field approach. The main difference between
the earlier scheme and this one is that the approximation will be applied to
the underlying \textit{dynamics} of the model \cite%
{LiuJoladSchZia13,LiuSchZia14} (as opposed to evaluating the sum $\Sigma
_{\left\{ \mathbb{N}\right\} }$ above). In other words, we formulate an
approximation on the \textit{transition probabilities} -- for the degree of
a particular node to increase/decrease by unity: $R\left( k\rightarrow k\pm
1\right) $. Once these are determined, we impose the steady state condition%
\begin{equation}
\tilde{\rho}\left( k\right) R\left( k\rightarrow k-1\right) =\tilde{\rho}%
\left( k-1\right) R\left( k-1\rightarrow k\right)  \label{RR-rho}
\end{equation}%
to find $\tilde{\rho}\left( k\right) $ (being an approximate $\rho $, again
denoted by a tilde) in closed form. The strategy is as follows. Exploiting
particle-hole symmetry, we will consider the two distributions, $\rho _{I}$
and $\zeta _{E}$, as well as two sets of rates, $R_{I,E}$. Each $R$ will
depend on an unknown parameter, representing the average degree of the
opposite community. From these, explicit expressions for $\tilde{\rho}_{I}$
and $\tilde{\zeta}_{E}$ can be obtained. In turn, the average degrees can be
computed and the unknown parameters can now be fixed through
self-consistency. In this spirit, we refer to this scheme as a SCMF
approximation, details of which can be found in Appendix C. Here, we simply
quote the results:

\begin{equation}
\tilde{\rho}_{I}\left( k_{I}\right) =\frac{\lambda ^{N_{E}-k_{I}}}{%
Z_{I}\left( N_{E}-k_{I}\right) !};~~\tilde{\zeta}_{E}\left( p_{E}\right) =%
\frac{\mu ^{N_{I}-p_{E}}}{Z_{E}\left( N_{I}-p_{E}\right) !}  \label{DD-tilde}
\end{equation}%
where $\lambda ,\mu $ are constants which can be obtained from $N_{I,E}$
alone and $Z$'s are normalization factors (Eqs. \ref{ZI},\ref{ZE}). Both are 
\textit{truncated} Poisson distributions, since $k_{I}\in \left[ 0,N_{E}%
\right] $ and $p_{E}\in \left[ 0,N_{I}\right] $. Instead of quoting $\lambda 
$ and $\mu $ from the SCMF calculation, we plot the full distributions
predicted by Eqs.~(\ref{rho-tilde},\ref{zeta-tilde}), shown as solid black
lines in Fig.~\ref{DD}(a). We should emphasize that \textit{no} fit
parameters have been introduced in this approach; the lines depend only on
the control parameters, $N_{I,E}$. It is clear that the agreement between
theory and simulation data is excellent for $N_{I}>N_{E}$. By symmetry, it
will also be quite good for cases with $N_{I}<N_{E}$. Indeed, disagreement
between theory and data is visibly detectable only in the tails of the
next-to-critical case, $\left( 101,99\right) $, a sign that correlations can
no longer be entirely neglected here. From these distributions, we easily
obtain $\left\langle X\right\rangle $ using Eq. (\ref{constraint}). Unlike
the results from the previous Section, all of these predictions fall within
statistical errors of the data. Since mean field schemes are not the start
of a systematic set of approximations, it is unclear why the approach here
is so much more successful at capturing the essentials of the model.


\begin{figure}[tbp]
\centering
\includegraphics[width=3.0in]{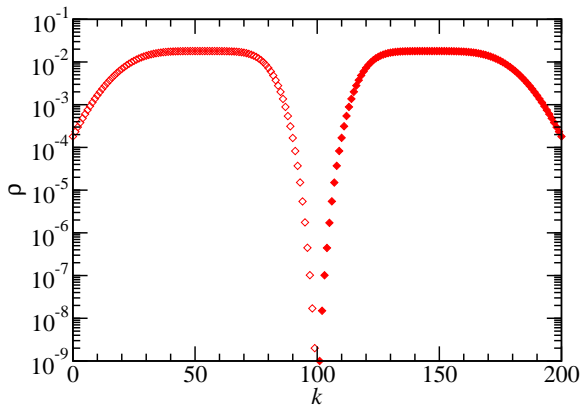}
\caption{(Color online) Simulation results of the degree distribution for the 
symmetric $\left( 100,100\right) $ case. The two components, denoted by open 
and solid diamonds, can be associated with the separate distributions 
$\protect\rho _{I}$ and $\protect\rho _{E}$, respectively.}
\label{100DD}
\end{figure}


To end this section, let us illustrate, with the symmetric case $(100,100)$,
the challenges of `criticality.' As expected, there are non-trivial
obstacles for both Monte Carlo simulations and theoretical understanding.
First of all, it takes much longer for the system to settle, typically a
hundred-fold longer than the $N_{I}\neq N_{E}$ cases. To compile a reliable
histogram for the $\rho \left( k\right) $, shown in Fig.~\ref{100DD}, we
take $10^{10}$ measurements in a combination of $5$ runs, each of which
lasts for $10^{11}$ MCS (after discarding the first $10^{7}$ MCS for the
system to settle into steady states). Note that, to untangle $\rho _{I,E}$
in the central pair of points, we recorded separately whether an $I$ or a $E$
node has $99$ or $100$ links. Of course, the distribution is symmetric,
i.e., $\rho _{I}=\zeta _{E}$. Unlike the off-critical cases, these
distributions display broad and flat plateaux. Undoubtedly, the physics
underlying these also gives rise to the plateau in $P\left( X\right) $.
Quantitative understanding of these large fluctuations remains elusive,
while the SCMF prediction for this case is, not surprisingly, far from ideal.

\section{Summary and outlook}

In this article, we report findings concerning an extraordinary phase transition
in a minimal model of dynamic social networks with preferred
degree. Consisting of \textit{extreme} introverts/extroverts, who only
cut/add connections, these dynamic networks have only the $I$-$E$ links and
reduce to bipartite graphs. With only two control parameters, $N_{I,E}$,
this seemingly trivial model displays surprising behavior. In particular, we
find compelling evidence that, in the limit of large populations, (i) the
likelihood of a link being present, $\left\langle x\right\rangle $, jumps
discontinuously, from $0$ to $1$, when $N_{I}$ drops below $N_{E}$ and (ii)
in the $N_{I}=N_{E}$ case, $x$ assumes all values in $\left[ 0,1\right] $
with equal probability. Such remarkable properties have been observed in
other statistical systems, e.g., 1D Ising models with certain long range
interactions \cite{BarMukamel14,BM14a}. We can place the similarity between 
our dynamic $XIE$ model and equilibrium Ising systems on firmer grounds, by
mapping their microscopic configurations one-to-one and regarding the
evolution of the former as Glauber spin-flip dynamics on the latter. Thanks
to restoration of detailed balance, we are able to find an exact expression
for the $XIE$ stationary distribution, $\mathcal{P}^{ss}$. Interpreting it
as a Boltzmann factor, $-\ln \mathcal{P}^{ss}$ can be regarded as a
`Hamiltonian' for the spins, though much more complex than typical Ising
models. Statistical properties of our system can now be explored along
standard routes.

To study of the collective behavior arising from such microscopics, we focus
on the degree distributions and $x$, the fraction of cross-connections. Since 
$x $ corresponds to $m$ (the magnetisation in the Ising model), while 
$h\,$or $\Delta $ (Eq. \ref{h+Delta}) can serve as an external magnetic
field, a natural question for us is: What is the `equation of state,' 
$m\left( h\right) $? Though naive expectations lead to the trivial $m=h$,
both simulations and mean field theories point towards the contrary: 
$m=sign\left( h\right) $, which is a hallmark of an extreme Thouless effect
\cite{BarMukamel14,BM14a}. Apart from the macroscopic $X$, we also study degree
distributions $\rho \left( k\right) $, `mesoscopic' quantities which are
commonly used in characterizing networks. Remarkably, the predictions from a
mean field approximation, formulated at the level of the underlying dynamics
for $\rho \left( k,t\right) $, are in excellent agreement with data (for all 
$N_{I}\neq N_{E}$).

The results of this first study are encouraging and provide us with good
stepping stones towards more systematic investigations. The most obvious
question may be what forms do thermodynamic limits take, assuming they
exist. Preliminary studies \cite{DharZia} suggest that, for fixed $h$, the
degree distributions (\ref{DD-tilde}) approach non-trivial limits. On the
other hand, if $\Delta $ were held fixed instead, it is unclear what the
limiting behavior is. How the system approach such limits is the next issue.
A detailed study of finite size scaling should be undertaken \cite{FGKB},
using both simulations and theoretical techniques. Pursuing an exact
computation of the partition function, and perhaps $P\left( X\right) $,
poses a worthy challenge. The failure of mean-field theory, especially near
`criticality' hints at the importance of correlations. However, there is no
spatial structure in our model and so, the usual notion of correlation
length is ill defined. Nevertheless, we have some evidence of strong
correlations, in the sense that the joint distribution, 
$\rho \left(k_{I},k_{E}\right) $, can be quite different from the product 
$\rho_{I}\left( k_{I}\right) \rho _{E}\left( k_{E}\right) $. Systematic
investigations of them are straightforward and worthwhile. From a theoretical
point of view, it would be desirable to develop an understanding for why the
SCMF is so much more successful than the standard mean-field approach
(Section III). Such insights may have impact beyond this study, as they may
reveal how best to formulate mean-field approximations.

Beyond exploring these questions, we can extend the $XIE$ model in an
orthogonal direction, arguably of purely theoretical interest (at present).
We may treat $\mathcal{H}$ as a genuine Hamiltonian in a standard study of
critical phenomena in thermal equilibrium. In other words, we propose to
study the statistical mechanics of a $L\times L$ system associated with the
Boltzmann factor%
\begin{equation}
\mathcal{P}\propto \exp \left\{ -\beta \left[ \mathcal{H}-BX\right] \right\} 
\end{equation}%
Here, $\beta $ is the usual inverse temperature variable, while the bias $B$
plays the role of a symmetry breaking, `magnetic field' (similar, but not
identical to $h$ in the $XIE$). It is interesting to note that, while the
critical control parameters of a typical system (e.g., $T_{c}$ in Ising, 
$T_{c}$ and $p_{c}$ for liquid gas) are not known, they are given precisely
by $\beta _{c}=1$ and $B_{c}=0$ here. For this `purely theoretical' system,
work is in progress \cite{FGKB}, to explore the usual avenues of
interest: static and dynamic critical exponents, scaling functions,
universality and the classes, etc. In the language of renormalization group
analyses (which proved to be highly effective in dealing with other
mixed-order transitions \cite{Cardy81,BarMukamel14}), we already know that 
$\mathcal{H}$ lies on the critical sheet and can inquire about fixed points
and their neighborhoods, irrelevant and relevant variables (e.g., if there
are others besides $\beta -1$ and $B$), etc. In addition, there are unusual
challenges, such as the lack of a natural correlation length in such a
system.

Beyond the $XIE$ model and its purely theoretical companion, there is a wide
vista involving dynamic networks with preferred degrees. For instance,
instead of assigning one or two $\kappa $'s to a population \cite%
{LiuJoladSchZia13,LiuSchZia14}, it is more natural to assign a distribution
of $\kappa $'s. There are also multiple ways to model interactions between
the various groups. For example, even with just two groups, it is realistic
to believe that an individual may have \textit{two }preferred degrees, one
for contacts within the group and another for those outside. Surely, this
kind of differential preference underlies the formation of social cliques.
Beyond understanding the topology and dynamics of interacting networks of
the types described here, the next natural step is to take into account the
freedom associated with the nodes, e.g., opinion, wealth, health, etc., on
the way to the ambitious goal of understanding adaptive, co-evolving,
interdependent networks in general. Along the way, we can expect the
unexpected, such as the emergence of the extreme Thouless effect in this $XIE
$ model, arguably the simplest of all interacting social networks.

\bigskip

\appendix

\section{Restoration of detailed balance}

In this appendix, we show that all Kolmogorov loops are reversible in the
XIE model and so, detailed balance is restored \cite{Kolmogorov36}. Since the
full dynamics occurs on the $\mathcal{N}$ cross-links, the configuration
space consists of the corners of an $\mathcal{N}$-dimensional unit cube,
while adding/cutting a link is associated with traversing an edge therein.
Clearly, products of the ratios of forward and reversed transition rates
around any closed loop can be expressed in terms of those around `elementary
loops' -- i.e., loops around a plaquette on the $\mathcal{N}$-cube. We will
show that the ratio associated with every plaquette is unity and so, all
Kolmogorov loops are reversible.

First, it is easy to see that if an elementary loop consists of modifying
two links connected to four different nodes, then the actions on each link
are unaffected by the other. In other words, rates associated with opposite
sides of the square (loop) are the same. Thus, their product in one
direction is necessarily the same as in the reverse. We need to focus only
on situations where the two links are connected to three nodes, e.g., $ij$
and $im$. For any such loop, let us start with a configuration in which both
are absent ($n_{ij}=n_{im}=0$). Let the states of node be such that $i$ has 
$k_{i}$ links, and $j$ and $m$ have $p_{j}$ and $p_{m}$ `holes',
respectively. Then one way around the loop is adding these two links
followed by cutting them, which can be denoted as the sequence 
\begin{equation}
\binom{n_{ij}}{n_{im}}=\binom{0}{0}\rightarrow \binom{1}{0}\rightarrow 
\binom{1}{1}\rightarrow \binom{0}{1}\rightarrow \binom{0}{0}
\end{equation}%
and leaving the rest of $\mathbb{N}$ unchanged. The associated product of
the transition rates is, apart from an overall factor of $N^{4}$, 
\begin{equation}
\frac{1}{p_{j}}\frac{1}{p_{m}}\frac{1}{k_{i}+2}\frac{1}{k_{i}+1}
\label{product}
\end{equation}%
Now, the reversed loop can be denoted as 
\begin{equation}
\binom{n_{ij}}{n_{im}}=\binom{0}{0}\rightarrow \binom{0}{1}\rightarrow 
\binom{1}{1}\rightarrow \binom{1}{0}\rightarrow \binom{0}{0}
\end{equation}%
associated with the product 
\begin{equation}
\frac{1}{p_{m}}\frac{1}{p_{j}}\frac{1}{k_{i}+2}\frac{1}{k_{i}+1}
\end{equation}%
which is exactly equal to Eqn.~(\ref{product}). From symmetry, we can expect
the same results for loops involving two introverts and one extrovert (i.e., 
$ij$ and $kj$). Thus, we conclude that the Kolmogorov criterion is satisfied
and detailed balance is restored in this XIE limit. Our system should settle
into a stationary distribution without probability currents, much like the
Boltzmann distribution for a system in thermal equilibrium.

\section{Considerations for computing $\Omega $ and $P\left( X\right) $}

Exploiting 
\begin{equation}
k!=\int_{0}^{\infty }due^{-u}u^{k}
\end{equation}%
the `partition function' can be expressed as%
\begin{equation*}
\Omega =\sum_{\left\{ n_{ij}\right\} }\prod\limits_{i}\int_{0}^{\infty
}du_{i}e^{-u_{i}}u_{i}^{\Sigma _{j}n_{ij}}\prod\limits_{j}\int_{0}^{\infty
}dv_{j}e^{-v_{j}}v_{j}^{\Sigma _{i}\bar{n}_{ij}}
\end{equation*}%
where we have used Eqn. (\ref{kpbar}). Exchanging the configuration sum and
integrals, we can perform the former to find%
\begin{equation*}
\Omega  = \int \mathcal{D}u\mathcal{D}ve^{-\left[ \Sigma _{i}u_{i}+\Sigma
_{j}v_{j}\right] }\sum_{\left\{ n_{ij}\right\}
}\prod\limits_{i,j}u_{i}^{n_{ij}}v_{j}^{\bar{n}_{ij}}  \label{PF1} 
\end{equation*}%
which can be cast as
\begin{equation*}
\int \mathcal{D}u\mathcal{D}v\exp \left\{ -\sum_{i,j}\left[ 
\frac{u_{i}}{N_{E}}+\frac{v_{j}}{N_{I}}+\ln \left( u_{i}+v_{j}\right) \right] \right\} 
\label{PF2}
\end{equation*}%
Here, 
\begin{equation}
\int \mathcal{D}u\mathcal{D}v\equiv \int_{0}^{\infty
}\prod\limits_{i}du_{i}\prod\limits_{j}dv_{j}
\end{equation}%
is a precursor for functional integrals if we take the continuum limit 
$u_{i},v_{j}\rightarrow u\left( x\right) ,v\left( y\right) $ and regard the
resultant as a two component, 1-D field theory. An attempt to use standard 
steepest descent leads to the following complications. As long as 
$N_{I}\neq N_{E}$, the maximum of the integrand in (\ref{PF2}) is located 
at a boundary of the region of integration.
However, for $N_{I}=N_{E}$, the maximum is a line given by $u_{i}=\bar{u}%
,v_{j}=\bar{v}$ and $\bar{u}+\bar{v}=N_{I,E}$. Both are non-standard
behavior and require more care to proceed.

Similar considerations can be given to the computation of $P\left( X\right) $,
in the sense that its generating function%
\begin{equation}
G\left( z\right) \equiv \sum_{X}z^{X}P\left( X\right)
\end{equation}%
is given by $\tilde{\Omega}\left( z\right) /\Omega $, where%
\begin{equation}
\tilde{\Omega}\left( z\right)  \equiv \sum_{\left\{ n_{ij}\right\}
}z^{X}\prod\limits_{i=1}^{N_{I}}\left( k_{i}!\right)
\prod\limits_{j=1}^{N_{E}}\left( p_{j}!\right)  
\end{equation}%
i.e.,
\begin{equation}
\int \mathcal{D}u\mathcal{D}v\exp \left\{ -\sum_{i,j}\left[ 
\frac{u_{i}}{N_{E}}+\frac{v_{j}}{N_{I}}+\ln \left( zu_{i}+v_{j}\right) \right] \right\} 
\end{equation}%
Of course, this integral is fraught with the same issues as in $\Omega $. 
Obviously, even if $G\left(z\right) $ can be found, inverting it may pose other 
challenges. Thus, deriving the presence of a plateau in $P\left( X\right) $, 
as well as anomalous exponents associated with its edges, will be a non-trivial
endeavour.

\section{Degree distributions in a self consistent mean-field approach}

In this appendix, we provide some technical details for the SCMF scheme.
Note that parameters ($\lambda ,\mu $) in the main text are, in fact, the
physically meaningful quantities ($\left\langle p_{E}\right\rangle ^{\prime
},\left\langle k_{I}\right\rangle ^{\prime }$) here.

First, consider a particular $I$ node, with $\rho _{I}\left( k_{I}\right) $
being the probability to find it having $k_{I}$ links. Then, provided $%
k_{I}>0$, $R_{I}\left( k_{I}\rightarrow k_{I}-1\right) =1/N$ which is the
probability that this node is chosen to act. By contrast, the exact rate for
having a link added ($k_{I}-1\rightarrow k_{I}$) is more complicated, since
it depends not only on all the $N_{E}-k_{I}+1$ extroverts \textit{not}
connected to it, but also on how many `holes' each has -- through $1/p_{j}$
(in Eq.~\ref{rates}). To proceed, we make judicious approximations. In the
spirit of mean-field theory, we can replace $1/p_{j}$ by the average $%
\left\langle 1/p_{E}\right\rangle ^{\prime }$, where the prime stands for an
average \textit{restricted} to nodes with $p_{E}>0$. Though we can formulate
the theory with $\left\langle 1/p_{E}\right\rangle ^{\prime }$, let us make
a further simplifying approximation and replace it by $1/\left\langle
p_{E}\right\rangle ^{\prime }$. So, we write%
\begin{equation}
R_{I}\left( k_{I}-1\rightarrow k_{I}\right) \cong \frac{N_{E}-k_{I}+1}{N}%
\frac{1}{\left\langle p_{E}\right\rangle ^{\prime }}  \label{RI}
\end{equation}%
So far, $\left\langle p_{E}\right\rangle ^{\prime }$ is an unknown
parameter. \textit{If} we had the distribution of an extrovert's holes, $%
\zeta _{E}\left( p_{E}\right) $, then we have the following relation:%
\begin{equation}
\left\langle p_{E}\right\rangle ^{\prime }\equiv \frac{\sum_{p_{E}>0}p_{E}%
\zeta _{E}\left( p_{E}\right) }{\sum_{p_{E}>0}\zeta _{E}\left( p_{E}\right) }%
=\frac{\langle p_{E}\rangle }{1-\zeta _{E}\left( 0\right) }  \label{p'}
\end{equation}%
But, $\zeta _{E}\left( p_{E}\right) $ is unknown. Nevertheless, at this
stage, we can exploit Eq.~(\ref{RR-rho}) and readily find 
\begin{eqnarray}
\tilde{\rho}_{I}\left( k_{I}\right) &=&\frac{N_{E}-k_{I}+1}{\left\langle
p_{E}\right\rangle ^{\prime }}\frac{N_{E}-k_{I}+2}{\left\langle
p_{E}\right\rangle ^{\prime }}...\frac{N_{E}}{\left\langle
p_{E}\right\rangle ^{\prime }}\tilde{\rho}_{I}\left( 0\right)  \notag \\
&\propto &\frac{\left( \left\langle p_{E}\right\rangle ^{\prime }\right)
^{N_{E}-k_{I}}}{\left( N_{E}-k_{I}\right) !}  \label{rhoI}
\end{eqnarray}

Before continuing to study $R_{E}$, let us work with this expression
further. Since $p_{I}\equiv N_{E}-k_{I}$ is the number of `holes' associated
with an $I$ node, we recognize this as a Poisson distribution (truncated at $%
N_{E}$) for the \textit{hole} distribution. Imposing normalization, we find
a compact closed form, $\tilde{\zeta}_{I}\left( p_{I}\right) =\left(
\left\langle p_{E}\right\rangle ^{\prime }\right) ^{p_{I}}/Z_{I}p_{I}!$,
where%
\begin{equation}
Z_{I}=\sum {}_{\ell =0}^{N_{E}}\left( \left\langle p\right\rangle ^{\prime
}\right) ^{\ell }/\ell !  \label{ZI}
\end{equation}%
is the sum of the first $N_{E}+1$ terms of an exponential series. Despite
its simplicity, the notation $\tilde{\zeta}_{I}\left( p_{I}\right) $ may be
too confusing and so, we will quote the final result for $\tilde{\rho}_{I}$
as 
\begin{equation}
\tilde{\rho}_{I}\left( k_{I}\right) =\frac{\left( \left\langle
p_{E}\right\rangle ^{\prime }\right) ^{N_{E}-k_{I}}}{Z_{I}\left(
N_{E}-k_{I}\right) !}  \label{rho-tilde}
\end{equation}%
with $\left\langle p_{E}\right\rangle ^{\prime }$ being a to-be-determined
parameter.

Next, we turn to a particular $E$ node and, exploiting `particle-hole'
symmetry, consider its hole distribution, $\zeta _{E}\left( p_{E}\right) $.
Since adding a link is decreasing $p_{E}$ by unity, we again have $%
R_{E}\left( p_{E}+1\rightarrow p_{E}\right) =1/N$, the probability that this
node is chosen to act, provided $p_{E}>0$. Meanwhile, it is connected to $%
N_{I}-p_{E}$ (i.e., $k_{E}-N_{E}+1$) introverts, each of which has $k_{i}$
links. As above, we rely on the same arguments and replace the $k_{i}$'s by
a suitable average:%
\begin{equation}
R_{E}\left( p_{E}\rightarrow p_{E}+1\right) \cong \frac{N_{I}-p_{E}}{N}\frac{%
1}{\left\langle k_{I}\right\rangle ^{\prime }}  \label{RE}
\end{equation}%
where 
\begin{equation}
\left\langle k_{I}\right\rangle ^{\prime }=\frac{\langle k_{I}\rangle }{%
1-\rho _{I}\left( 0\right) }  \label{k'}
\end{equation}%
Recasting Eq.~(\ref{RR-rho}) for $\tilde{\zeta}$, we have 
\begin{equation}
\tilde{\zeta}_{E}\left( p_{E}\right) =\frac{\left\langle k_{I}\right\rangle
^{\prime }}{N_{I}-p_{E}}\tilde{\zeta}_{E}\left( p_{E}+1\right)
\end{equation}%
Again, this recursion relation leads to a (truncated) Poisson distribution
in $N_{I}-p_{E}$, and imposing normalization, we have explicitly 
\begin{equation}
\tilde{\zeta}_{E}\left( p_{E}\right) =\frac{\left( \left\langle
k_{I}\right\rangle ^{\prime }\right) ^{N_{I}-p_{E}}}{Z_{E}\left(
N_{I}-p_{E}\right) !}  \label{zeta-tilde}
\end{equation}%
with 
\begin{equation}
Z_{E}=\sum {}_{\ell =0}^{N_{I}}\left( \left\langle k\right\rangle ^{\prime
}\right) ^{\ell }/\ell !  \label{ZE}
\end{equation}%
Of course, $\left\langle k_{I}\right\rangle ^{\prime }$ here is also an
unknown, to-be-determined, parameter. Note that, along with Eq.~(\ref%
{rho-tilde}), this result again confirms the underlying particle-hole
symmetry.

Finally, we make the last approximation. Instead of the exact (and unknown)
parameters, $\left\langle p_{E}\right\rangle ^{\prime }$ and $\left\langle
k_{I}\right\rangle ^{\prime }$, let us approximate them by using $\tilde{\rho%
}_{I}$ and $\tilde{\zeta}_{E}$ in Eqs.~(\ref{k'},\ref{p'}) instead. Since 
$\tilde{\rho}_{I}$ and $\tilde{\zeta}_{E}$ depend on $\left\langle
p_{E}\right\rangle ^{\prime }$
and $\left\langle k_{I}\right\rangle ^{\prime} $, respectively, 
we may define the functions $f$ and $g$:%
\begin{eqnarray}
\left\langle k_{I}\right\rangle ^{\prime } &\cong& 
\frac{\Sigma k_{I}\tilde{\rho}_{I}\left( k_{I}\right) }{1-\tilde{\rho}_{I}\left( 0\right) }
\equiv f\left(\left\langle p_{E}\right\rangle ^{\prime }\right) \\
\left\langle p_{E}\right\rangle ^{\prime } &\cong& 
\frac{\Sigma p_{E}\tilde{\zeta}_{E}\left(p_{E}\right) }{1-\tilde{\zeta}_{E}\left( 0\right) }
\equiv g\left(\left\langle k_{I}\right\rangle ^{\prime }\right)
\end{eqnarray}%
Making a plot of these functions in the $\left\langle k\right\rangle
^{\prime }$-$\left\langle p_{E}\right\rangle ^{\prime }$ plane, the point of
intersection then determines, self-consistently, the values for these two
parameters. In practice, it is simple to start with, say, a trial value 
$p_{0}$ for $\left\langle p_{E}\right\rangle ^{\prime }$ and compute 
$\left\langle k_{I}\right\rangle ^{\prime }$ through Eq.~(\ref{rho-tilde}).
Inserting this $\left\langle k_{I}\right\rangle ^{\prime }$ into 
Eqn.~(\ref{zeta-tilde}), we compute $\tilde{\zeta}_{E}$ and the associated 
$\left\langle p_{E}\right\rangle ^{\prime }$. If this result is not $p_{0}$,
then vary the latter until they agree. In other words, this process will
lead us to the solution: $\left\langle p_{E}\right\rangle ^{\prime }=g\left(
f\left( \left\langle p_{E}\right\rangle ^{\prime }\right) \right) $.
Substituting these values ($\left\langle p_{E}\right\rangle ^{\prime }$ and 
$\left\langle k_{I}\right\rangle ^{\prime }$) into 
Eqs.~(\ref{rho-tilde},\ref{zeta-tilde}), the degree distributions can be plotted.

\bigskip

\begin{acknowledgements}
Illuminating discussions with A. Bar, J. Cardy, D. Dhar, Y. Kafri, W. Kob, S. Majumdar, 
D. Mukamel, and Z. Toroczkai are gratefully acknowledged. We thank C. del Genio and 
F. Greil for valuable technical advice. This research is supported in part
by the US National Science Foundation, through grants DMR-1206839 (KEB) and
DMR-1244666 (WL, BS, and RKPZ), 
and by the AFOSR and DARPA through grant FA9550-12-1-0405 (KEB).
One of us (RKPZ) thanks the Galileo Galilei Institute for Theoretical
Physics for hospitality and the INFI for partial support during the
completion of this paper.
 \end{acknowledgements}

\bibliography{XIE}
\end{document}